# Resilience of the physicochemical properties of graphene-based materials for applications in harsh radiation environments

**M.A. Guazzelli[a], S.G. Alberton[b], N. Added[b], V.A.P. Aguiar[b], K. Araki[c], L.H. Avanzi[a], F. Cappuzzello[d,e], M. Cavallaro[e], E.F. Chinaglia[a], M.T. Escote[f], F.F. Ferreira[f], M. Giovannini[g], R.F. Jardim[b], S.H. Masunaga[a,b], N.H. Medina[b], M. Nakamura[c], J.R.B. Oliveira[b], R.B.B. Santos[a], A.C. Vilas-Bôas[a]**

[a] Centro Universitário da FEI, São Bernardo do Campo, São Paulo, Brazil
[b] Instituto de Física, Universidade de São Paulo, São Paulo, Brazil
[c] Instituto de Química da Universidade de São Paulo, São Paulo, SP, Brazil
[d] Dipartimento di Fisica e Astronomia "Ettore Majorana", Università di Catania, Catania, Italy
[e] Istituto Nazionale di Fisica Nucleare–Laboratori Nazionali del Sud, Catania, Italy
[f] Universidade Federal do ABC, Santo André, São Paulo, Brazil
[g] University of Genoa, Italy

**Abstract**. The development of radiation-tolerant materials capable of maintaining structural, electrical, and thermal stability in extreme, radiation-rich environments remains a critical challenge in materials science. In this work, the effects of 60 MeV $^{35}$Cl ion irradiation on highly oriented pyrolytic graphite (HOPG) and multilayer reduced graphene oxide (ML-rGO) were investigated. The samples were exposed to fluences of $5.11 \times 10^9$ and $1.3 \times 10^{10}$ ions/cm² and characterized by X-ray diffraction (XRD), Raman spectroscopy, scanning electron microscopy (SEM), atomic force microscopy (AFM), and electrical transport measurements. The results show that the irradiation response is strongly influenced by the initial structural organization of the material. In HOPG, ion exposure leads to a progressive loss of crystalline order, evidenced by XRD peak broadening and an increase in the Raman ID/IG ratio, accompanied by a reduction in electrical transport performance. In contrast, ML-rGO exhibits distinct behavior at higher fluences, suggesting partial structural reorganization. The appearance of more defined graphitic features in XRD and Raman analyses, along with changes in surface morphology and electrical response, suggests the formation of more ordered sp² domains. These findings indicate that irradiation effects vary with the initial degree of order, providing useful insights for selecting carbon-based materials for devices operating under severe radiation conditions.

## 1. Introduction

The development of radiation-tolerant materials is a critical challenge for advancing applications in extreme, radiation-rich environments, such as those found in particle accelerators, aerospace engineering, and advanced nuclear physics experiments [1-7]. A major concern in these contexts is the substantial heat generated during intense heavy-ion irradiation in nuclear reactions, which can cause melting and premature degradation of thin-film targets and substrates, thereby compromising long-term experimental reliability [1, 6-9]. Consequently, there is a critical need to develop substrate materials that can withstand extreme thermal and structural stresses while efficiently dissipating lateral heat. Highly Oriented Pyrolytic Graphite (HOPG) has been widely studied and proposed as a backing material for such demanding applications [1, 5, 8, 9]. Composed of stacked, highly ordered graphene layers, HOPG exhibits exceptional in-plane thermal conductivity and mechanical integrity [10, 11]. However, the thermal and electrical performance of HOPG is strongly dependent on its crystalline order, which is highly sensitive to radiation-induced defects [1, 5, 12-15]. Previous studies have demonstrated that exposure to ionizing radiation leads to rapid accumulation of vacancies and interstitials, and to bond rearrangements in HOPG, causing a transition toward an amorphous carbon phase and significantly degrading the material's functional properties [1, 15-26].

Multilayer reduced graphene oxide (ML-rGO) has emerged as a promising alternative due to its restored π-conjugated carbon network, high intrinsic thermal conductivity, and structural flexibility. Unlike HOPG, ML-rGO exhibits turbostratic stacking with varying degrees of pre-existing structural disorder, depending on its synthesis [10-13, 27-30]. The response of ML-rGO to ionizing radiation is complex and heavily fluence-dependent. While low-energy ion bombardment can induce severe structural disorder and reduce electrical conductivity by opening a bandgap, the specific structural evolution of ML-rGO under high-energy heavy-ion beams remains underexplored [1, 5, 16].

Existing studies on carbon-based backings have primarily focused on the predictable degradation trajectories of highly crystalline materials, leaving a gap in understanding how pre-existing disorder influences radiation tolerance [10, 14-27, 31-33]. To address this, the present study investigates the comparative effects of 60 MeV $^{35}$Cl ion beam irradiation on the structural, morphological, and electrical properties of HOPG and ML-rGO.

We hypothesize that the initial microstructural order fundamentally dictates the radiation response trajectory of these graphene-based multilayer materials. The key novelty of this work lies in demonstrating a profound dichotomy in material behavior: while heavy-ion irradiation systematically degrades the highly ordered lattice of HOPG, it paradoxically acts as an annealing agent for the disordered ML-rGO, driving an atomic-scale structural reorganization and partial recovery of crystallinity at specific fluences. By elucidating the relationship between irradiation-induced structural effects and alterations in physicochemical properties, this research establishes critical criteria for selecting and deploying radiation-tolerant materials in extreme environments [27, 31, 34].

## 2. Materials and methods

### *2.1. Materials*

Highly Oriented Pyrolytic Graphite (HOPG) and multilayer reduced graphene oxide (ML-rGO) samples were obtained commercially from Optigraph GmbH, and NanoTechnology Solutions, respectively. The HOPG samples had a thickness of (2.0 ± 0.5) µm, while the ML-rGO samples exhibited a thickness of (3.3 ± 0.4) µm. Both materials had lateral dimensions of approximately (1.0 × 1.0) $cm^2$ and were used as received for the subsequent irradiation and characterization procedures.

### *2.2 Methods*

#### *2.2.1. Ion beam irradiation*

HOPG and ML-rGO samples were irradiated with a 60 MeV $^{35}Cl$ ion beam in order to evaluate their response under extreme radiation-rich environments. The choice of the 60 MeV $^{35}Cl$ ions represented a compromise between achieving high ion flux, elevated linear energy transfer (LET), and significant vacancy creation, while simultaneously ensuring that the ions traversed both irradiated foils with nearly constant LET and vacancy generation values throughout the samples, as predicted by SRIM simulations [5]. The irradiation was performed at the Pelletron 8UD accelerator of the Laboratório Aberto de Física Nuclear e Aplicações (LAFNA, Instituto de Física da Universidade de São Paulo, Brazil) using the SAFIIRA beamline [35]. The samples were exposed to two distinct fluences: N1 = $5.1 \times 10^9$ and N2 = $1.3 \times 10^{10}$ particles/$cm^2$, with a mean flux of approximately $4.2 \times 10^5$ particles/s/$cm^2$. To ensure homogeneous and equivalent exposure, the samples were mounted on a goniometer that enabled continuous rotational control, allowing the positional exchange between the two samples every 20 minutes. As illustrated in Fig. 1, the samples were fixed on opposite sides of the goniometer, enabling alternating exposure during rotation. The beam was monitored using solid-state detectors and scattered by gold foils to reduce flux, increase the interaction area, and improve homogeneity.

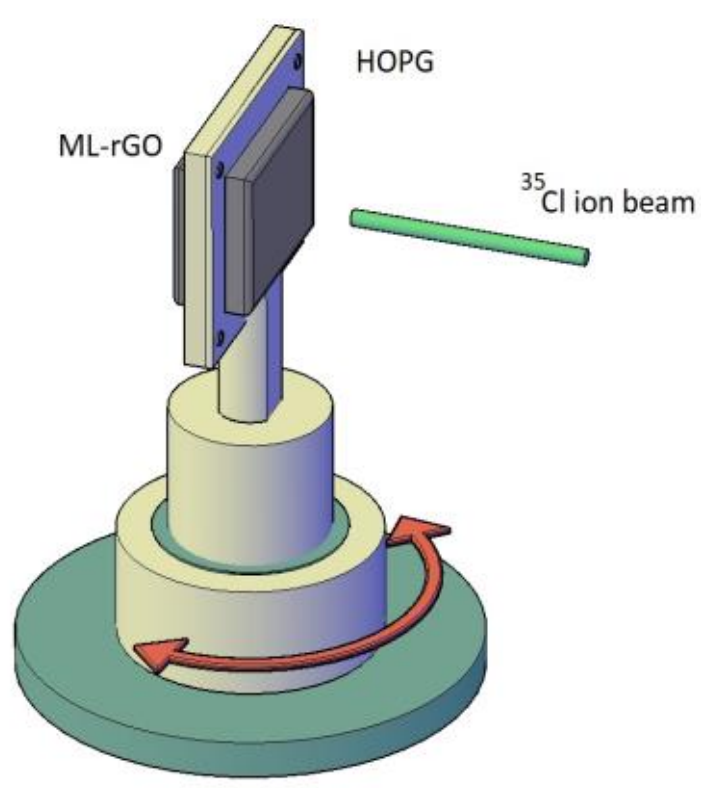

**Fig. 1.** Schematic of the experimental setup in the SAFIIRA beamline, with ML-rGO and HOPG samples mounted on a rotating goniometer. The configuration allows periodic swapping during 60 MeV $^{35}$Cl ion irradiation, ensuring uniform exposure.

After each irradiation step (N1 and N2), the samples were sectioned into smaller pieces to enable the different characterization techniques to be performed on representative portions of the same material. In total, two HOPG and two ML-rGO samples were irradiated with 60 MeV $^{35}$Cl ions, with one sample of each material sequentially exposed to both fluence levels. This approach ensured that all analyses were performed on equivalent regions under identical irradiation conditions, minimizing sample-to-sample variability and enabling consistent comparisons across the different techniques.

### *2.2.2. Material characterization*

To assess the structural and morphological modifications induced by the ion beam, both pristine and irradiated samples were characterized using the following techniques:

- X-ray Diffraction (XRD): High-resolution XRD data were collected using a STADI-P diffractometer (Stoe®, Darmstadt, Germany) operating in transmission geometry at 40 kV and 40 mA. The measurement was performed with Cu $K\alpha_1$ radiation ($\lambda$ = 1.54056 Å) filtered by a primary-beam Ge(111) curved-crystal monochromator, a 0.5 mm divergence slit, a 3 mm circular scattering slit, and a silicon strip detector (Mythen 1K, Dectris®, Baden, Switzerland). Scans were performed in the 2θ angular range of 25.000º to 93.250º, with a step size of 0.015º and an integration time of 180 s at each 1.05º [1].
- Raman Spectroscopy: Micro-Raman spectral images were acquired using a WITEC alpha 300R confocal Raman microscope equipped with an Nd:YAG laser ($\lambda$ = 532 nm). Measurements were conducted utilizing a Nikon objective (100×, NA = 0.8) and a laser power of 30 mW/cm$^2$ on the sample stage, with a beam diameter of 400 nm. Raman mapping was performed over (10 × 10) µm$^2$ areas (35 × 35) pixels, corresponding to a 286 nm pixel size) with an integration time of 1 s per pixel [1].

- Scanning Electron Microscopy (SEM): Surface microstructural changes and nanoscale defects were visually assessed using a Field Emission Scanning Electron Microscope (FEG-SEM, Tescan Mira 3 XMU) operating at an accelerating voltage of 30 kV with 1.0 nm resolution [1].
- Atomic Force Microscopy (AFM): Topographical imaging and surface roughness (RMS) analysis were conducted in air, in dynamic mode, using a Shimadzu® SPM9700 microscope with commercial Si tips. Scans were acquired at various sizes ranging from 1.0 µm to 8.0 µm to evaluate average roughness and step-through distances [1].

#### *2.2.3. Electrical transport measurements*

Temperature-dependent electrical resistance measurements were performed using a Physical Property Measurement System (PPMS, Quantum Design) over a temperature range from 10 K to 300 K [1]. The measurements were conducted using the standard four-contact method, with electrical connections established using gold wires attached to the sample surface with silver paint [36].

## 3. Results and discussion

### 3.2. Structural evolution under heavy ion irradiation

#### 3.2.2. X-ray diffraction (XRD) analysis

Figure 2 shows the X-ray diffraction (XRD) patterns, obtained in transmission geometry, for all samples. It is important to emphasize that the pristine and N1 ML-rGO samples shown in Fig. 2(b) do not exhibit well-defined diffraction peaks.

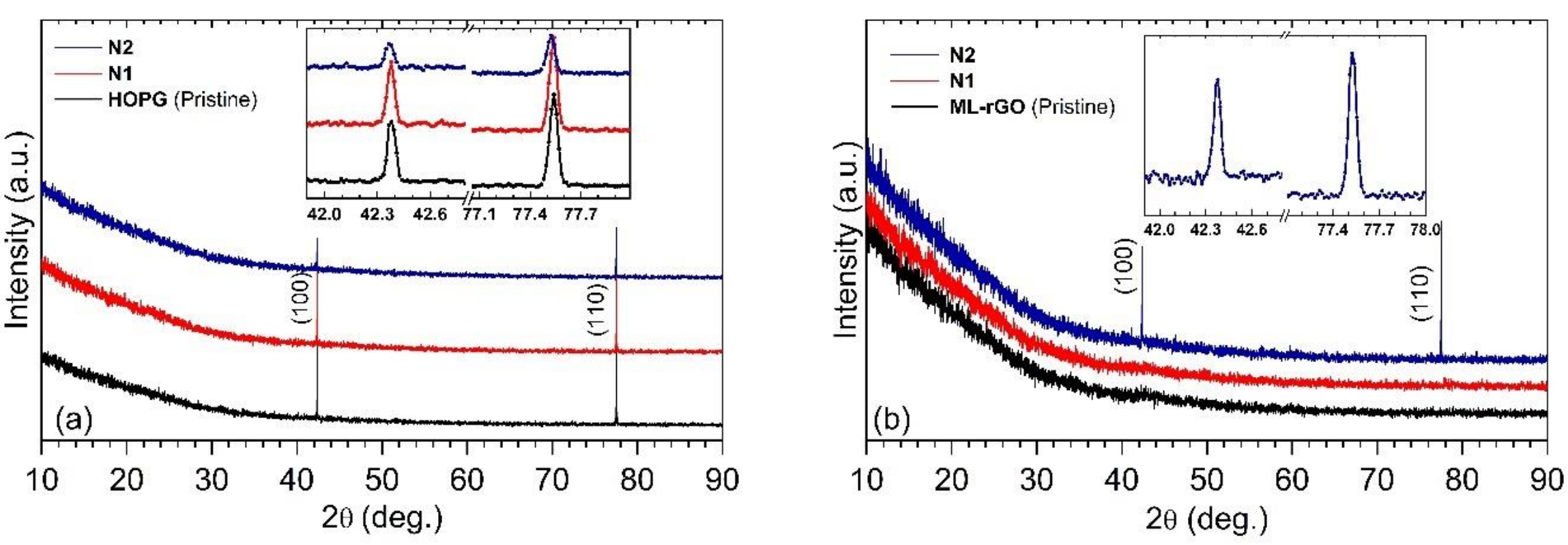

**Fig. 2.** X-ray diffraction (θ–2θ) patterns of the analyzed samples: (a) HOPG; and (b) ML-rGO, both shown for pristine and after exposure to a $^{35}Cl$ ion beam, highlighting the effects of irradiation on the structural properties of the materials.

For HOPG, a clear decreasing trend in diffraction peak intensities is observed as the irradiation fluence increases, accompanied by subtle changes in the lattice parameters and an increase in the full width at half maximum (FWHM), as summarized in Table 1. The *a* parameter fluctuates from 2.46072 Å (pristine) to 2.46103 Å (N1) and 2.46060 Å (N2), while the FWHM increases from 0.0509° to 0.0597°. This attenuation in peak intensity is attributed to the generation of point defects and the disruption of long-range crystalline order via kinetic-energy transfer mechanisms, characteristic of atomic displacements and vacancy formation [1, 28, 30, 37].

**Table 1**
FWHM and lattice parameters variation for all HOPG samples

| Sample | FWHM (º) | *a* (Å) | $\Delta a/a$ (%) |
|---|---|---|---|
| Pristine | 0.0509(4) | 2.46072(2) | - |
| N1 | 0.0544(14) | 2.46103(18) | +0.013(7) |
| N2 | 0.0597(11) | 2.46060(13) | -0.005(5) |

In contrast, the ML-rGO samples exhibit a distinct structural trajectory. The pristine and N1 ($5.1 \times 10^{9}$ ions/cm²) samples exhibit an amorphous halo characteristic of the turbostratic or disordered stacking typical of reduced graphene oxide. However, after irradiation at the higher fluence (N2 = $1.3 \times 10^{10}$ ions/cm²), characteristic diffraction peaks similar to those of highly ordered HOPG emerge, indicating a remarkable ion-beam-induced ordering of the crystal structure [11]. This recrystallization is likely driven by the localized removal of oxygen-containing functional groups that act as layer spacers, facilitating the restacking of graphene sheets from a disordered amorphous state into a more ordered graphitic structure [29].

### 3.2.3. Raman spectroscopy

Micro-Raman spectroscopy further elucidates the divergent structural responses of these materials.

As shown in Fig. 3a and 3b, the pristine HOPG spectrum exhibits an intense G band and a negligible D band, confirming a high-quality sp² graphitic network. After irradiation, subtle modifications are observed, including the appearance of a 2D band at ~2706 cm⁻¹ and a monotonic increase in the ID/IG ratio (from 0.0040 to 0.0065), indicating the progressive introduction of point defects, such as vacancies and interstitials, and the partial disruption of hexagonal lattice symmetry [1, 38-40].

In contrast, the ML-rGO samples shown in Fig. 3c and Fig. 3d undergo significant structural evolution. While the pristine and N1 spectra display broad features characteristic of disordered carbon, the N2 condition reveals a sharpening of the G band at ~1576 cm⁻¹, the reappearance of a well-defined 2D band at ~2706 cm⁻¹, and a noticeable reduction in the ID/IG ratio. These changes, along with the ID/IG ratios summarized in Table 2, suggest a partial restoration of graphitic ordering. The energy deposited by ion irradiation likely induces transient local heating via the thermal spike mechanism, promoting defect annealing and the coalescence of graphitic domains [1, 29, 38, 41, 42].

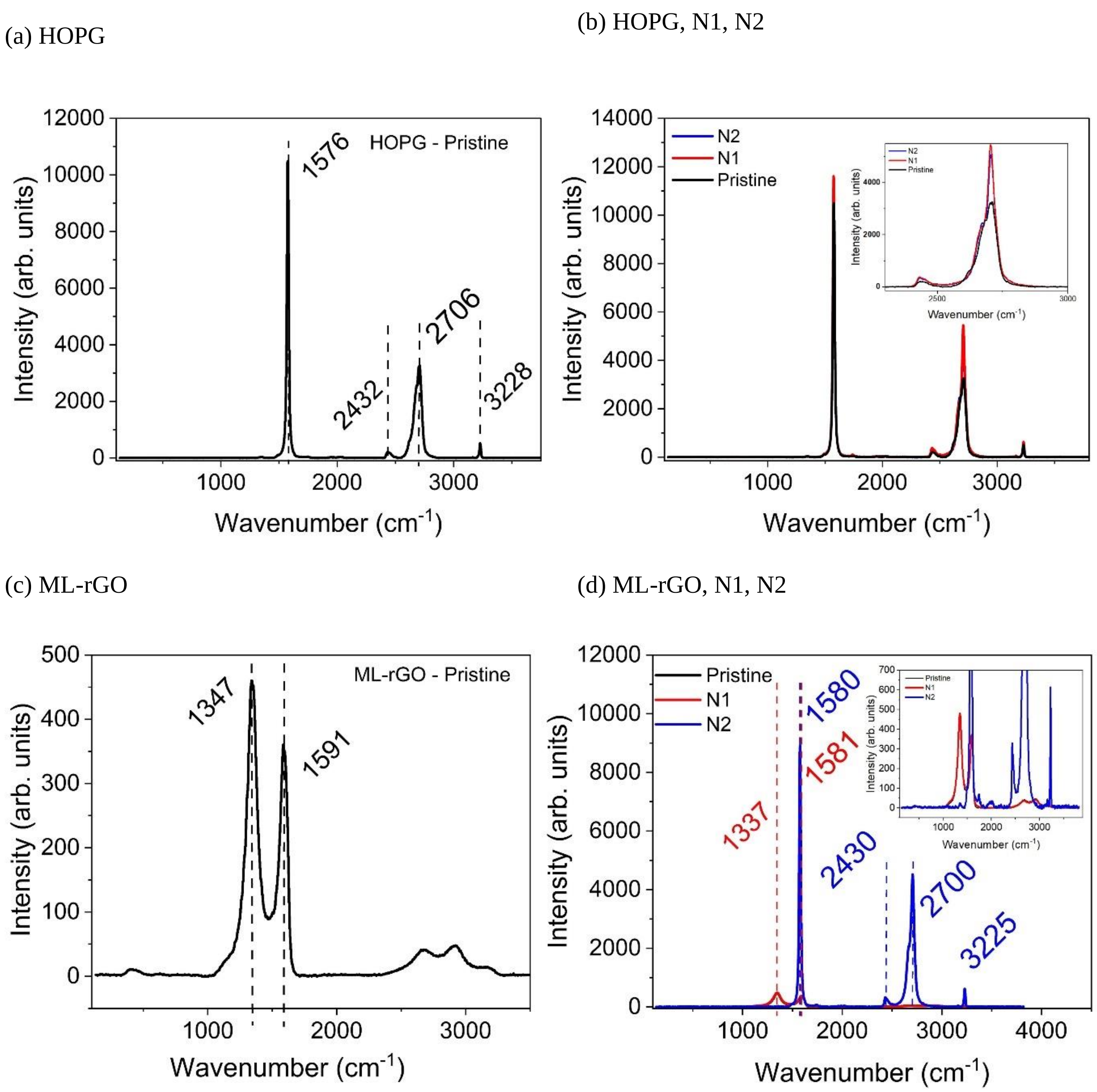

**Fig. 3**. Micro-Raman spectra of the analyzed samples: (a) pristine HOPG, highlighting the characteristic Raman bands associated with its high structural order; (b) HOPG under pristine and irradiated conditions (N1 and N2), enabling comparison of irradiation effects; (c) pristine ML-rGO, showing the typical defect-related D and G bands; and (d) ML-rGO for pristine and irradiated conditions (N1 and N2). The spectra in (b) and (d) correspond to the average Raman response from selected regions in the Raman maps where the D band is present, allowing assessment of irradiation-induced disorder.

**Table 2**
D/G disorder relation for HOPG and ML-rGO foil

| Sample | ID/IG | ID/IG |
| --- | --- | --- |

| | HOPG | ML-rGO |
|---|---|---|
| Pristine | 0.0040 (10) | 1.22 (6) |
| N1 | 0.0039 (10) | 1.29 (7) |
| N2 | 0.0065 (7) | 0.0076 (11) |

The micro-Raman mapping results presented in Fig. 4, based on the spatial distribution of the D-to-G band intensity ratio (I_D/I_G), reveal a clear evolution of the structural properties of ML-rGO as a function of irradiation fluence. In pristine and N1 conditions, the maps exhibit a relatively homogeneous distribution with high ID/IG values, indicating that the D band is uniformly present and more intense than the G band across the mapped area, consistent with the turbostratic and defect-rich nature of ML-rGO [11 – 29]. In contrast, under the N2 condition, a pronounced reduction in the ID/IG ratio is observed, with values approaching 0–0.01 a.u., indicating that the D band is largely suppressed over most of the mapped region. This behavior reflects a significant decrease in defect density and the emergence of more ordered graphitic domains. To better visualize the remaining defective regions, a binarized image was generated, highlighting only the pixels where the D band is still present. A comparison with the binarized map of pristine HOPG, which exhibits highly ordered domains, shows that the ML-rGO sample after N2 irradiation develops isolated micro-domains with similar spatial characteristics. These results indicate that high-fluence irradiation promotes substantial structural reorganization, leading to graphitic regions with a degree of crystallinity comparable to that of pristine HOPG, in agreement with the Raman spectral analysis.

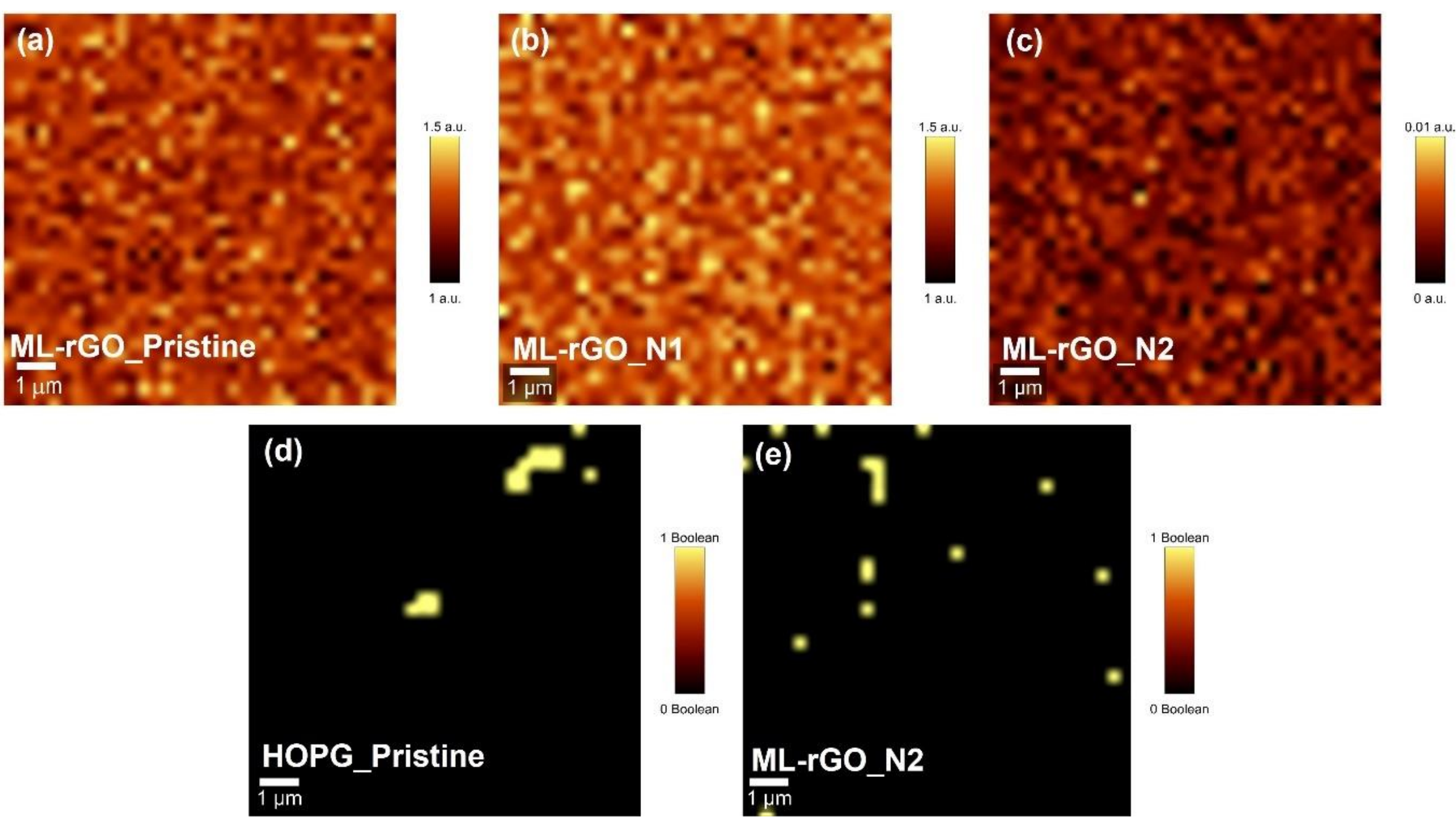

**Fig. 4.** Micro-Raman maps (10 × 10 μm²) of ML-rGO under different irradiation conditions: (a) pristine sample, (b) after the first irradiation step (N1), and (c) after the second irradiation step (N2). Binarized images are shown for (d) pristine HOPG, and (e) ML-rGO after N2.

### 3.3. Surface morphology and microstructural analysis

As shown in Fig. 5, scanning electron microscopy (SEM) images reveal pronounced morphological differences between the materials, corroborating the spectroscopic results.

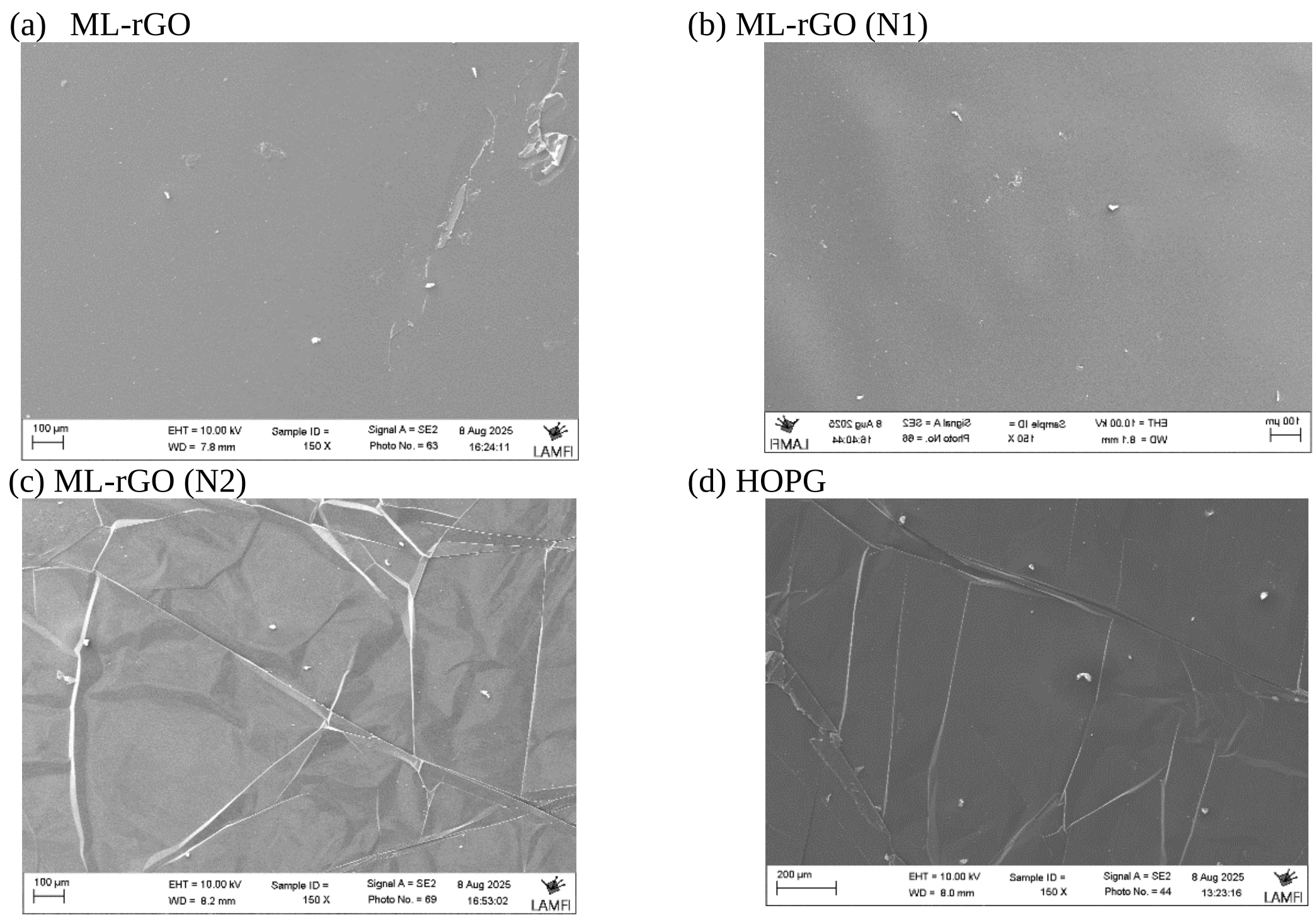

**Fig. 5.** Scanning electron microscopy (SEM) images of the analyzed samples: (a) ML-rGO pristine, (b) ML-rGO after the first irradiation (N1), (c) ML-rGO after the second irradiation (N2), and (d) HOPG pristine. The pristine samples highlight the pronounced structural differences between ML-rGO and HOPG, where ML-rGO exhibits a more homogeneous and irregular morphology characteristic of partial disorder, while HOPG shows well-defined planes associated with ordered graphene layer stacking. The irradiated ML-rGO samples (N1 and N2) evidence the progressive morphological changes induced by ion exposure.

As shown in Fig. 5a and 5b, the pristine ML-rGO and N1 samples exhibit relatively smooth and featureless surfaces, consistent with a partially disordered, turbostratic-like structure. Following the second irradiation step (N2), Fig. 5c reveals the development of more defined planar features, indicating a partial structural reorganization toward lamellar domains resembling graphitic layer stacking. This morphological evolution is in good agreement with the Raman results (Fig. 3), which show a reduction in the ID/IG ratio (Table 2) and the reappearance of a well-defined 2D band, as well as with the XRD data, which suggest improved structural ordering [11, 43]. In contrast, the HOPG reference sample (Fig. 5d), exposed to the same irradiation conditions, exhibits clear signs of irradiation-induced degradation, including increased surface roughness associated with atomic displacements and sputtering processes, as well as indications of out-of-plane expansion, consistent with lattice swelling effects. These observations are consistent with the Raman analysis, which indicates an increase in disorder (higher ID/IG ratio), and with the XRD results, which show peak broadening (increased FWHM, Table 1) and a reduction in diffraction intensity, reflecting the progressive disruption of long-range crystalline order.

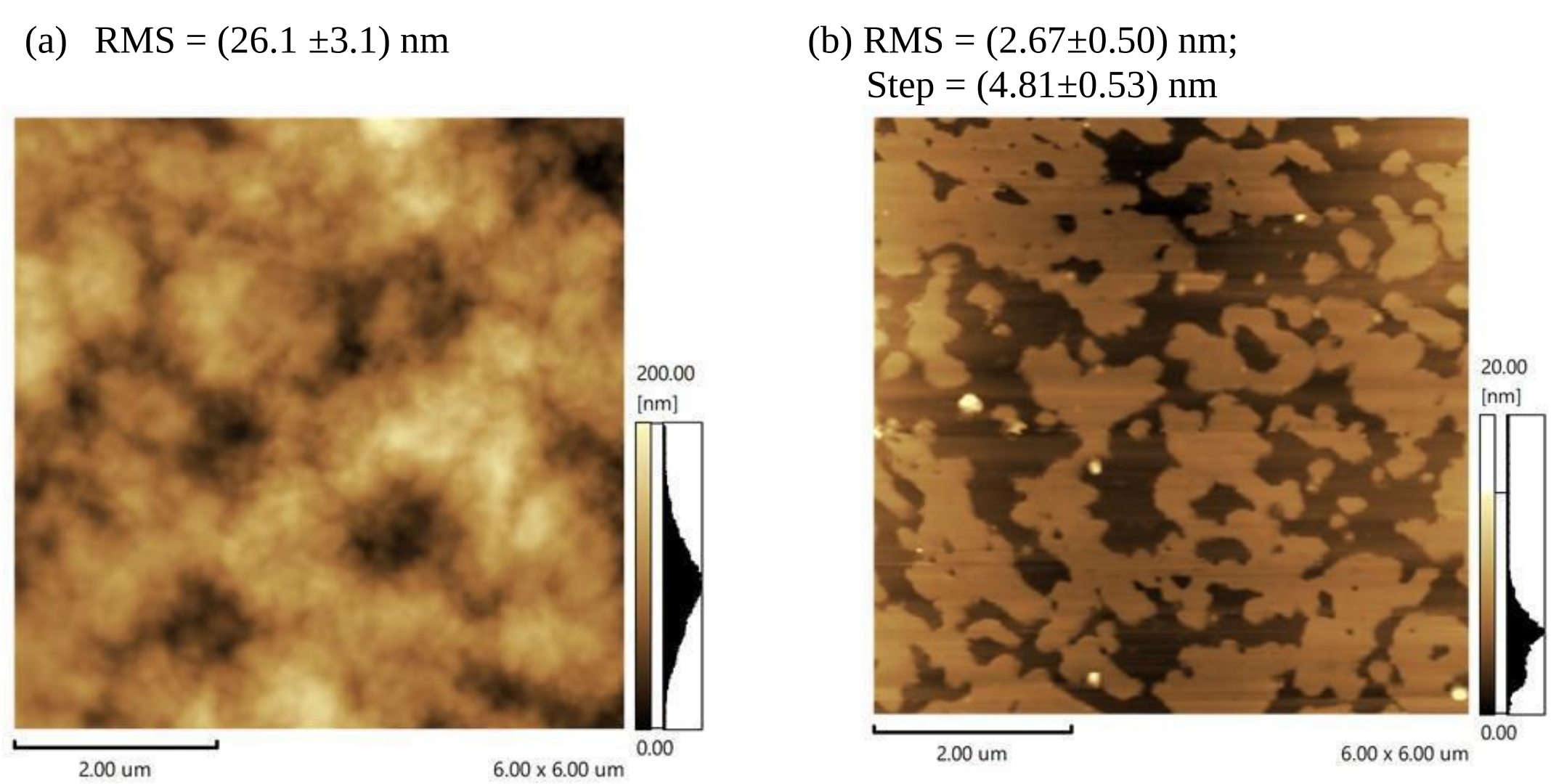

**Fig. 6.** Atomic force microscopy (AFM) images of ML-rGO samples: (a) pristine, exhibiting high surface roughness characteristic of a partially disordered structure; (b) after the second irradiation (N2), showing the emergence of planar features with step-like structures and reduced surface roughness, indicative of structural reorganization toward more ordered domains.

Atomic force microscopy (AFM) analysis, shown in Fig. 6, provides quantitative support for the observed morphological evolution [1]. As presented in Fig. 6a, the pristine ML-rGO sample exhibits a highly irregular surface, with a root-mean-square (RMS) roughness of (26.1 ± 3.1) nm, consistent with residual structural disorder and randomly stacked flakes. After the second irradiation step (N2), Fig. 6b reveals the formation of well-defined planar regions, characterized by an average step height of (4.81 ± 0.53) nm and a markedly reduced RMS roughness of (2.67 ± 0.50) nm.

This pronounced reduction in surface roughness, together with the emergence of step-like features, supports the interpretation that controlled ion irradiation promotes structural reorganization, driving the transition from a highly disordered morphology toward a more ordered, graphene-like arrangement in ML-rGO materials [11, 28, 29, 38].

## 3.4. Electrical transport properties

As shown in Fig. 7, the temperature-dependent electrical resistivity was analyzed to assess the microstructural evolution and functional properties, with particular emphasis on the Residual Resistivity Ratio (RRR = R(300 K)/R(10 K)).

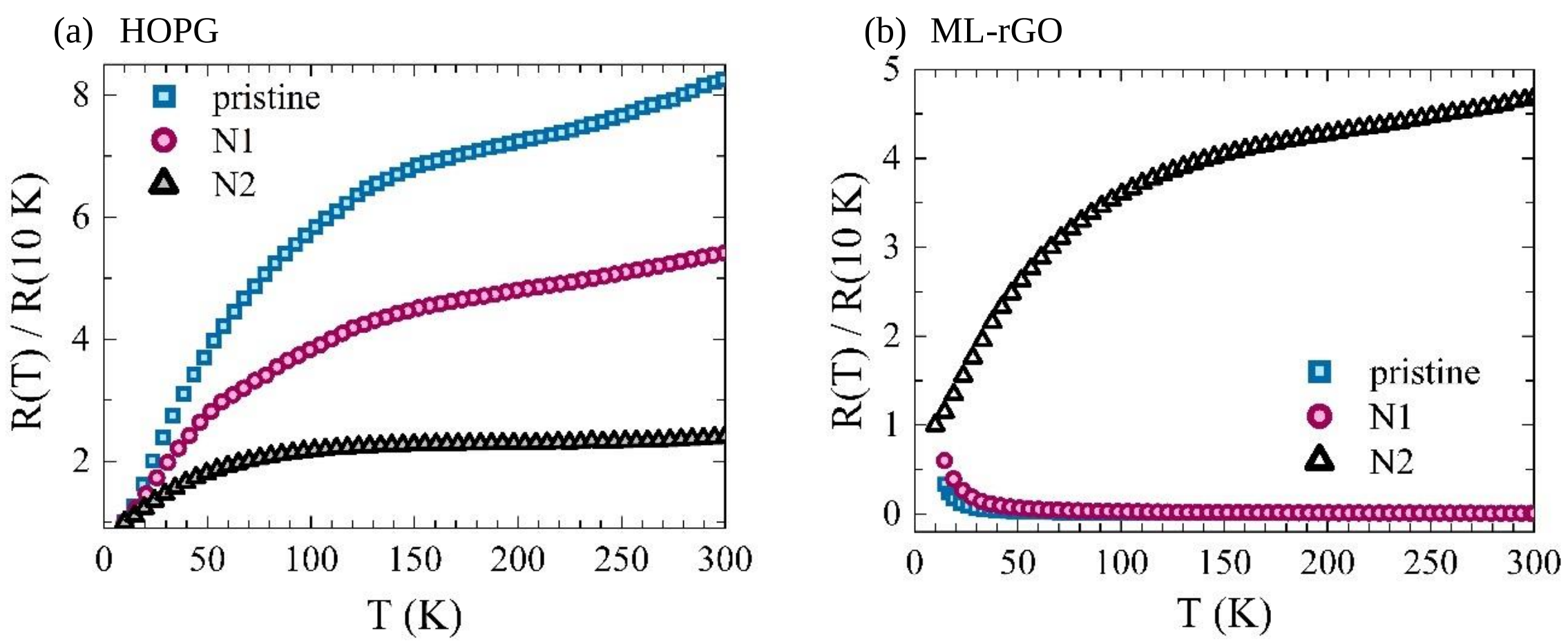

**Fig. 7.** Temperature dependence of the normalized resistance, R(T)/R(10 K), for (a) HOPG and (b) ML-rGO samples, shown for the pristine, N1, and N2 specimens.

**Table 3**
Residual resistivity ratio (RRR) values for HOPG and ML-rGO samples in the pristine and irradiated (N1 and N2) conditions.

| Sample | RRR HOPG | RRR ML-rGO |
|---|---|---|
| Pristine | 8.3(4) | - |
| N1 | 5.4(3) | - |
| N2 | 2.40(12) | 4.7(2) |

As shown in Fig. 7a, all HOPG samples exhibit a positive temperature coefficient of resistance ($d\rho/dT > 0$, where $\rho$ is the electrical resistivity of the material), consistent with the semimetallic behavior of graphite. However, a systematic reduction in the Residual Resistivity Ratio (RRR) is observed with increasing irradiation fluence, as summarized in Table 3, decreasing from (8.3±0.4) (pristine) to (5.4±0.3) (N1) and (2.40±0.12) (N2). This progressive suppression of RRR indicates the degradation of long-range crystalline order and the accumulation of irradiation-induced point defects, which enhance carrier scattering, particularly at low temperatures. In contrast, the ML-rGO samples, shown in Fig. 7b, display a non-monotonic evolution of their transport properties. The pristine and N1 samples exhibit a negative temperature coefficient of resistance ($d\rho/dT < 0$), characteristic of semiconducting behavior in disordered, turbostratic carbon networks, where charge transport is limited by defect-induced potential barriers. A marked transition is observed for the N2 sample, which exhibits a positive temperature coefficient ($d\rho/dT > 0$) and an RRR value of (4.7±0.2) (Table 3), indicating a shift toward metallic-like conduction. This behavior suggests that ion irradiation at higher fluence induces partial structural recovery, likely driven by thermal spike effects, promoting defect annealing and the formation of more continuous conductive pathways, consistent with a transition toward a polycrystalline graphitic structure.

### 3.4. Comprehensive correlation of structural, morphological, and electrical properties

The multi-technique characterization provides a mechanistic understanding of how ion-induced structural modifications dictate the electrical transport properties of the investigated carbon materials. Rather than isolated phenomena, morphological and structural alterations directly govern charge-carrier mobility across the irradiated matrices.

For the highly crystalline HOPG, the progressive attenuation of crystalline order, evidenced by the broadening of XRD reflections (Fig. 2, Table 1) and the rise in the Raman ID/IG ratio (Fig. 3 and 4, Table 2), directly correlates with the severe degradation of its electrical performance. The kinetic energy transfer from the 60 MeV $^{35}Cl$ ions generates a high density of point defects (vacancies and interstitials). Morphologically, this atomic disruption induces lattice swelling along the c-axis and increased surface roughness (Fig. 5 and 6). In terms of transport kinetics, these structural defects act as potent scattering centers that hinder charge carrier mobility at low temperatures. Consequently, the mean free path of the electrons is drastically reduced, which fundamentally accounts for the 71% suppression of the Residual Resistivity Ratio (RRR) observed in the N2 sample (Fig.7, Table 3).

Conversely, the correlative analysis of the ML-rGO samples reveals a thermal-spike-induced annealing mechanism that facilitates the formation of conductive networks. Initially, the turbostratic structure and residual oxygen functionalities limit charge transport through defect-mediated hopping, resulting in semiconductor-like behavior (Fig. 7). However, the localized energy deposition at the higher fluence (N2) provides sufficient activation energy to heal the $sp^2$ carbon network. This microstructural recovery is confirmed by the emergence of graphitic XRD peaks and the sharpening of the Raman 2D band. Topographically, the transition manifests as a reduction in AFM surface roughness to 2.67 nm and the formation of well-defined planar terraces (Fig.6, b). This ion-beam-induced planarization is critical for the electrical transport: the coalescence of ordered graphitic domains removes the defect-induced potential barriers and establishes continuous, low-resistance percolative pathways. This structural ordering effectively explains the macroscopic transition from a defect-dominated semiconducting state to a metallic-like conduction regime (Table 3).

## Conclusion

The present results show that the response of carbon-based materials to heavy-ion irradiation is strongly governed by their initial structural organization. Under 60 MeV $^{35}$Cl ion exposure, HOPG and ML-rGO follow markedly different evolution pathways, reflecting the role of crystalline order in defining radiation tolerance.

- In HOPG, irradiation induces a progressive loss of structural order. The reduction in XRD peak intensity, the broadening of XRD peaks, and the increase in the Raman ID/IG ratio consistently indicate the accumulation of lattice defects. This degradation is accompanied by a continuous decrease in the residual resistivity ratio, evidencing the direct impact of defect generation on electrical transport.
- ML-rGO exhibits a distinct behavior, particularly at higher fluences. Instead of a purely degradative process, irradiation promotes partial structural reorganization. The appearance of graphitic features in XRD patterns, together with the sharpening of Raman bands and reduction of the ID/IG ratio, suggests a tendency toward increased ordering.
- Morphological analyses support this interpretation, showing the development of more planar structures and a reduction in surface roughness. These changes are reflected in the electrical response, with a transition from semiconducting to more metallic-like conduction, consistent with the formation of more extended $sp^2$ domains.
- Despite this apparent improvement, the evolution of ML-rGO is not monotonic and shows significant variability with fluence. This lack of control over irradiation-induced modifications can compromise the reproducibility and stability required for device applications.

From an application standpoint, the two materials present complementary limitations and advantages. HOPG undergoes continuous degradation, but in a predictable manner, which is beneficial for reliability assessment and lifetime estimation in radiation environments. In contrast, ML-rGO shows irradiation-induced structural rearrangements that may locally enhance its properties, yet without a consistent or controllable trend.

These findings are particularly relevant for the development of electronic devices and components intended to operate under severe radiation conditions, where both performance stability and predictability are critical. The results emphasize that material selection should account not only for intrinsic properties but also for how these properties evolve under irradiation. Such considerations are essential for the design of robust systems in radiation-demanding environments.

**Acknowledgments**

The authors acknowledge financial support from the funding agencies: INCT-FNA proc. n. 408419/2024-5; CNPq: proc. n. 301576/2022-0, proc. n. 404054/2023-4, proc. n. 303295/2022-8, proc. n. 408800/2021-6, proc. N. 316019/2021-6, proc. n. 44259/2019-6 and n.304651/2021-4, FAPESP proc. n. 2023/16053-8, proc. n. 2024/00989-7 and proc. n. 2024/01816-9, SisNANO/MCTI, Centro Universitário FEI and the NUMEN Project.